\documentclass[prb,amsmath,superscriptaddress,showpacs,showkeys,twocolumn]{revtex4-1}
\usepackage{amssymb,amsmath}   
\usepackage[dvips]{graphicx}   
\usepackage{verbatim}   
\usepackage{color}      
\usepackage{subfigure}  
\usepackage{hyperref}   
\usepackage{gensymb}
\usepackage{epstopdf}
\usepackage{natbib}
\usepackage{enumerate}
\DeclareMathOperator{\sinc}{sinc}
\begin{document}

\title{Decoherence of a quantum two-level system by spectral diffusion}
\author{Shlomi Matityahu}
\affiliation{Department of Physics, Ben-Gurion University of the
Negev, Beer Sheva 84105, Israel}
\author{Alexander Shnirman}
\affiliation{Institut f\"ur Theorie der Kondensierten Materie,
Karlsruhe Institute of Technology, 76131 Karlsruhe, Germany} \affiliation{L. D. Landau Institute
for Theoretical Physics RAS, Kosygina street 2, 119334 Moscow,
Russia}
\author{Gerd Sch\"on}
\affiliation{Institut f\"ur Theoretische Festk\"orperphysik,
Karlsruhe Institute of Technology, 76131 Karlsruhe, Germany}
\affiliation{Institute of Nanotechnology, 
Karlsruhe Institute of Technology, D-76344 Eggenstein-Leopoldshafen, Germany}
\author{Moshe Schechter}
\affiliation{Department of Physics, Ben-Gurion University of the
Negev, Beer Sheva 84105, Israel}
\date{\today}
\begin{abstract}
We study the dephasing of an individual high-frequency tunneling two-level system (TLS) due
to its interaction with an ensemble of low-frequency thermal TLSs which are described 
by the standard tunneling model (STM). 
We show that the dephasing by the bath of TLSs explains
both the dependence of the Ramsey dephasing rate on an externally
applied strain as well as its order of magnitude, as observed in
a recent experiment [J. Lisenfeld {\it et al.}]. However, the
theory based on the STM predicts the Hahn-echo protocol to be much more
efficient, yielding too low echo dephasing rates, as compared to the experiment. 
Also the strain dependence of
the echo dephasing rate predicted by the STM does not agree with the measured
quadratic dependence, which would fit to a high-frequency white
noise environment. We suggest that few fast TLSs which are coupled much more strongly to the strain
fields than the usual TLSs of the STM give rise to such a white noise. This explains the magnitude and strong fluctuations of the echo dephasing rate observed in the experiment.
\end{abstract}

 \maketitle
\section{Introduction} \label{Introduction} The low-temperature
physics of amorphous and disordered solids has been a subject of
great interest for more than four decades. Below about $1\,$K the
acoustic and thermodynamic properties of a large variety of
glasses are not only qualitatively different compared to their
crystalline counterparts, but show a remarkable degree of
universality.\cite{ZRC71,BJF88,PRO02} This universal behavior was
explained by the existence of low-energy excitations with
two-level structure, known as tunneling two-level systems
(TLSs).\cite{APW72,PWA72} In the standard tunneling model (STM),
each TLS is characterized by the energy bias $\Delta$ between its two basis
states, the tunnel splitting $\Delta^{}_{0}$ between both, and its coupling $\gamma$ to the strain
field $\epsilon$. Thus the Hamiltonian of each TLS in the STM
is\cite{APW72,PWA72,PWA87}
\begin{align}
\label{eq:STM}&\hat{\mathcal{H}}^{}_{\mathrm{STM}}=\frac{1}{2}\left(\Delta\hat{\tau}^{}_{z}+\Delta^{}_{0}\hat{\tau}^{}_{x}\right)+\gamma
\epsilon\,\hat{\tau}^{}_{z}\, ,
\end{align}
where $\hat{\tau}^{}_{x}$ and $\hat{\tau}^{}_{z}$ are the Pauli
matrices  in the local states basis. The central idea of the STM is that the TLSs
are independent and their energy bias and tunnel splitting are
randomly distributed, with a universal distribution function
$f(\Delta,\Delta^{}_{0})=P^{}_{0}/\Delta^{}_{0}$, where $P^{}_{0}$
is a material dependent constant.\cite{PWA87}

The role of the coupling to the strain field [third term in Eq.~(\ref{eq:STM})] is threefold. 
First, via this channel the TLS couples to phonons, which gives rise to the energy 
relaxation of the TLS. Second, by applying mechanical stress externally, one can control and 
manipulate the TLS.\cite{GGJ12,LJ15}  Third, this coupling is at least partially responsible 
for the coupling between the TLSs.  Although the STM has originally considered independent TLSs, it is
now well established that TLSs interact via elastic
(phonon-mediated) or electric (photon-mediated) dipole
interactions,\cite{BJL77,BAL98,SM08} described by the low-energy
effective Hamiltonian
\begin{align}
\label{eq:int1}&\hat{\mathcal{H}}^{}_{\mathrm{int}}=\frac{1}{2}\sum^{}_{i\neq
j}J^{}_{ij}\hat{\tau}^{}_{z,i}\hat{\tau}^{}_{z,j}\, ,
\end{align}
where $\hat{\tau}^{}_{z,i}$ and $\hat{\tau}^{}_{z,j}$ are the
pauli matrices that represent the TLSs at sites $i$ and $j$, and
the interaction coefficients are
\begin{align}
\label{eq:int2}&J^{}_{ij}=\frac{c^{}_{ij}U^{}_{ij}}{R^{3}_{ij}}\,
.
\end{align}
The parameter $c^{}_{ij}\sim O(1)$ in general is a complicated function of
the polar angle between the TLSs at sites $i$ and $j$, but frequently can
be treated as a normal-distributed random variable, $R^{}_{ij}$
is the distance between the TLSs, and $U^{}_{ij}$ characterizes
the specific interaction. Here we will focus on elastic
interactions for which
\begin{align}
\label{eq:int3}&U^{}_{ij}=\frac{\gamma^{}_{i}\gamma^{}_{j}}{\rho
v^{2}}\, ,
\end{align}
where $\rho$ and $v$ are the mass density and sound velocity,
respectively. Our results can be straightforwardly generalized to
the case where electric interactions are present.


With the advent of superconducting quantum bits (qubits),
the investigation of the low-temperature properties of amorphous
solids has gained further interest\cite{PE14} for two main reasons. First,
ensembles of TLSs are believed to be a major source of fluctuations leading to energy
relaxation and decoherence in superconducting qubits and microwave
resonators. Specifically, the coherence times of superconducting
qubits incorporating Josephson junctions were found to be strongly
affected by the presence of TLSs located in the tunnel barriers of
the junctions \cite{SRW04,MJM05} which are typically made of
amorphous aluminum oxide with a thickness of 2-3 nm. A deeper
understanding of the nature of TLSs and their role as sources of
decoherence is necessary in order to achieve high-fidelity
operation of superconducting qubits. Second, it was established
that superconducting phase qubits can be utilized to study the properties of
individual TLSs.\cite{SRW04,NM08} The qubit spectrum exhibits an
avoided level crossing at bias values for which the qubit is in
resonance with a certain TLS. In addition this resonant coupling allows one to
manipulate the quantum state of the TLS directly by coherent
single-pulse resonant driving \cite{LJ10} as well as to study the longitudinal 
($T^{}_{1}$) and transverse ($T^{}_{2}$)
relaxation times of individual TLSs.\cite{SY10,LJ10}

The individual TLSs that are being probed and manipulated by 
qubits are characterized by the level splitting of the order of that of the 
qubit ($\sim 2\pi \cdot 10$GHz), which is substantially higher than the 
 temperature in the experiment. To distinguish the (high-frequency, ''probed'') 
TLS from, e.g., the TLSs with energy splittings smaller or of the order of the temperature (thermal TLSs), 
we introduce for the former the subscript $p$ and rewrite its Hamiltonian (\ref{eq:STM}) as\cite{Strain} 
\begin{align}
\label{eq:ProbedH}&\hat{\mathcal{H}}^{}_{p}=\frac{1}{2}\left(\Delta_p(\epsilon_p)\hat{\tau}^{}_{z,p}+\Delta^{}_{0,p}\hat{\tau}^{}_{x,p}\right)\, .
\end{align}
The bias $\Delta_p$ is controlled 
by the applied strain $\epsilon_p$ as written in Eq.~(\ref{eq:STM}), i.e.\ 
$\Delta_p(\epsilon_p) = const + \gamma_p \epsilon_p$.
Recently, a static strain tuning of the energy bias of individual
TLSs in a superconducting phase qubit was
demonstrated.\cite{GGJ12,LJ15} The results support the assumption
of the STM that the coupling of TLSs to strain fields occurs
mainly through a change of the asymmetry bias $\Delta_p$, whereas
changes of $\Delta^{}_{0,p}$ are negligible. 

In Ref.\ \onlinecite{LJ15} the strain
dependence of the relaxation and coherence times of individual high-frequency TLSs was measured,
revealing the following interesting features:
\begin{enumerate}[(1)]
\item The pure dephasing rate $\Gamma^{}_{\varphi,R}$ of the probed TLS, as extracted
from the Ramsey (free induction) protocol, is linear with respect to
the asymmetry bias $|\Delta^{}_{p}|$ in the
vicinity of the symmetry point $\Delta^{}_{p}=0$. Outside the
window $|\Delta^{}_{p}|\lesssim 2\pi\cdot 1\,$GHz around the
symmetry point,\cite{Comment1} the Ramsey dephasing rate deviates
from the linear behavior, exhibiting small changes of slope or
irregularities.

\item The pure dephasing rate $\Gamma^{}_{\varphi,E}$, as extracted
from the Hahn-echo protocol, is much smaller than the Ramsey dephasing
rate (with ${\Gamma^{}_{\varphi,E}/\Gamma^{}_{\varphi,R}\lesssim 0.1}$~at
${\Delta^{}_{p}\approx 2\pi\cdot 1}\,$GHz), demonstrating the
efficiency of the echo protocol in filtering noise at frequencies
${\omega\ll 2\pi\cdot 1}\,{\rm MHz}\sim \Gamma^{}_{\varphi,R}$. Furthermore, the echo dephasing
rate grows quadratically with $\Delta^{}_{p}$ in the whole
measured range of $\Delta^{}_{p}$. This implies the presence of a
white-noise environment inducing energy fluctuations on a
time scale $\lesssim 1\,\mu\rm{s}$.
\end{enumerate}

The starting point of the theoretical description is the following:
The individual high-frequency probed TLS with energy splitting 
$E_p\equiv\sqrt{\Delta_p^2 + \Delta_{0,p}^2} \gg k_{\rm B}T$ suffers from decoherence
due to its interaction with an ensemble of thermal TLSs with energy splittings
lower than the temperature. Such thermal TLSs undergo random transitions (telegraph noise)
between their two eigenstates, thereby causing fluctuations in the
energy splitting $E_p$ of the probed high-frequency TLS. 
This is the mechanism of spectral diffusion
observed in ''hole-burning'' experiments.~\cite{AW75,AW78}
As we will show in the following this picture needs to be refined in several directions.

In this paper we provide a detailed theoretical analysis of the coherence properties of an
individual  high-frequency TLS coupled to an ensemble of thermal TLSs. 
The qubit used to probe serves merely as a tool and is not part of the modeling.
After providing a general 
theoretical description of the spectral diffusion we analyze the 
decoherence caused by an ensemble of thermal TLSs with 
parameters taken from the STM (we call these TLSs "$\tau$-TLSs"). 
A short description of this theory
was provided in Ref.\ \onlinecite{LJ15}. We show that by assuming
elastic dipole-dipole interactions as in
Eqs.~(\ref{eq:int1})-(\ref{eq:int3}), with a bath of thermal $\tau$-TLSs,
one may explain very well both the qualitative dependence on
strain as well as the order of magnitude of the Ramsey dephasing
rate. This demonstrates that the TLSs introduced in the STM form
the main low-frequency noise source in superconducting phase
qubits. Moreover, the suggested theory predicts the temperature
dependence of the Ramsey dephasing rate (at a fixed strain). For
the echo dephasing rate, on the other hand, the theory at this level predicts
that the contribution of $\tau$-TLSs is orders of magnitude smaller than the
experimental result. In addition, the strain dependence of the echo dephasing rate predicted
by the theory is in disagreement with the experiment. 

We further show that the experimental results on the echo dephasing rate 
could be potentially explained by the presence of a few fast thermal TLSs.\cite{ZankerQP}
These TLSs are characterized by parameters very different from those of the STM ($\tau$-TLSs). 
In particular they have a much stronger coupling to the strain, which leads 
to the enhanced relaxation (switching) rate. From the point of view of the probed TLS 
the fast thermal fluctuators produce Gaussian noise, since their switching rate 
is larger than their coupling to the probed TLS. This noise is white at relevant 
frequencies (frequencies lower than the typical switching rate of the fast thermal TLSs). 
This explains the quadratic dependence of the echo 
dephasing rate on the applied strain. Such strongly interacting TLSs (henceforth
denoted as $S$-TLSs)
were recently introduced within a
model suggesting a possible explanation for the universality of
acoustic and thermodynamic properties in disordered systems at low
temperatures.\cite{SM13} We estimate the order of magnitude for
the contribution of the $S$-TLSs to the echo dephasing rate and
discuss its relevance in accounting for the experimental findings.
We emphasize that in all our considerations we assume the probed
TLS to be a standard ($\tau$-)TLS, since these are the prevalent
TLSs at low temperatures.

The paper is organized as follows: in Sec.\ \ref{Sec 1} we first
define the spin-fluctuator model,\cite{PE02,GYM04,BJ09} and then
describe its results for the Ramsey and echo decay signals of the
probed TLS in different regimes of relaxation rates and TLS-TLS
couplings (Sec.\ \ref{Sec 1A}). We then discuss the important
distinction between average and typical results, which arises in
situations in which a quantity is dominated by contributions from
a small number of fluctuators (Sec.\ \ref{Sec 1B}). The model is
applied in Sec.\ \ref{Sec 2} to the case of an environment
consisting of weakly interacting TLSs as postulated in the STM. We
first describe the characteristic relaxation rates and TLS-TLS
couplings of such TLSs (Sec.\ \ref{Sec 2A}), and then derive the
results for the Ramsey (Sec.\ \ref{Sec 2B}) and echo (Sec.\
\ref{Sec 2C}) dephasing rates. 
In Sec.\ \ref{Sec 3} we generalize the model to the model with two types of TLSs
introduced in Ref.\ \onlinecite{SM13}. We first review the two-TLSs
model (Sec.\ \ref{Sec 3A}), estimate the characteristic relaxation
rates and TLS-TLS couplings of strongly interacting $S$-TLSs
(Sec.\ \ref{Sec 3B}), and then study the contribution of $S$-TLSs
to the echo dephasing rate (Sec.\ \ref{Sec 3C}). Our conclusions
are discussed and summarized in Sec.\ \ref{Sec 5}.
\section{The model}
\label{Sec 1}
\subsection{Decoherence}
\label{Sec 1A} We consider a single high-frequency probed TLS interacting
with a set of thermal TLSs via the Hamiltonian (\ref{eq:int1}). 
Thus, the Hamiltonian reads
\begin{align}
\label{eq:Hpj}\hat{\mathcal{H}}&=\frac{1}{2}\left(\Delta_p \hat{\tau}_{z,p} + \Delta_{0,p}\hat{\tau}_{x,p}\right) + \frac{1}{2}
\sum^{}_{j}\left(\Delta_j \hat{\tau}_{z,j} + \Delta_{0,j}\hat{\tau}_{x,j}\right) 
\nonumber\\&+\sum_j J_j \hat{\tau}_{z,p}\hat{\tau}_{z,j}
+\hat{\mathcal{H}}^{}_{bath}\, .
\end{align}
The first term of Eq.~(\ref{eq:Hpj}) describes the probed TLS. The second term describes the set of thermal TLSs. 
The coupling constants $J^{}_{j}$ denote the
coupling strength~(\ref{eq:int2}) between the thermal TLS at site $j$ and
the probed TLS. The term $\hat{\mathcal{H}}^{}_{bath}$ 
describes the internal interactions between the thermal TLSs,
as well as their coupling to phonons.

In the eigenbasis of the TLSs, the system is described by the Hamiltonian
\begin{align}
\label{eq:3}\hat{\mathcal{H}}&=\frac{1}{2}E^{}_{p}\hat{\sigma}^{}_{z}+\frac{1}{2}\sum^{}_{j}E^{}_{j}\hat{\alpha}^{}_{z,j}+\frac{1}{2}\hat{\mathcal{X}}\left(\hat{\sigma}^{}_{z}\cos\theta^{}_{p}-\hat{\sigma}^{}_{x}\sin\theta^{}_{p}\right)\nonumber\\
&+\hat{\mathcal{H}}^{}_{bath}\, .
\end{align}
Here $\hat{\sigma}^{}_{z}$ are the Pauli matrices in the
eigenbasis of the probed TLS and $\hat{\alpha}^{}_{z,j}$ are the
Pauli matrices in the eigenbasis of the thermal TLS at site $j$. The first
term of Eq.~(\ref{eq:3}) describes the probed TLS with energy
$E^{}_{p}=\sqrt{\Delta^{2}_{p}+\Delta^{2}_{0,p}}$ and
$\cos\theta^{}_{p}=\Delta^{}_{p}/E^{}_{p}$,
$\sin\theta^{}_{p}=\Delta^{}_{0,p}/E^{}_{p}$. 
All these are strain-dependent via $\Delta_p(\epsilon_p)$.
The second term
describes the set of thermal TLSs with energies
$E^{}_{j}=\sqrt{\Delta^{2}_{j}+\Delta^{2}_{0,j}} \lesssim T$, and the third
term describes the interactions between the probed TLS and the set
of thermal TLSs, with the operator $\hat{\mathcal{X}}$ given by
\begin{align}
\label{eq:4}&\hat{\mathcal{X}}=2\sum^{}_{j}J^{}_{j}\left(\hat{\alpha}^{}_{z,j}\cos\theta^{}_{j}-\hat{\alpha}^{}_{x,j}\sin\theta^{}_{j}\right)\,
.
\end{align}
Here $\cos\theta^{}_{j}=\Delta^{}_{j}/E^{}_{j}$ and
$\sin\theta^{}_{j}=\Delta^{}_{0,j}/E^{}_{j}$. 

The operator $\hat{\mathcal{X}}$ fluctuates since each thermal
TLS, i.e.\ TLS for which $E^{}_{j}<T$, switches randomly between
its two eigenstates. Such random transitions are driven by the
emission or absorption of phonons,\cite{PWA72,JJ72} or by internal
interactions between the TLSs.\cite{BAL98,BAL94} The fluctuations
in $\hat{\mathcal{X}}$ add a random contribution to the bare
energy $E^{}_{p}$ of the probed TLS through the longitudinal
coupling ($\propto\hat{\sigma}^{}_{z}$) in the third term of
Eq.~(\ref{eq:3}). This results in a dephasing of the probed TLS.
Henceforth, we will simplify the model by making the following
assumptions:
\begin{enumerate}[(1)]
\item We consider random telegraph noise of thermally excited TLSs
due to relaxation ($T^{}_{1}$) processes. TLSs with energy larger
than the temperature are frozen in their ground state and are thus
irrelevant for the dephasing of the probed TLS. We will thus
restrict the sum in Eq.~(\ref{eq:4}) to thermal TLSs. Further,
since the probed TLS and the thermal TLSs are non-resonant, we can
neglect the transverse term ($\propto\hat{\alpha}^{}_{x,j}$) of
$\hat{\mathcal{X}}$ in Eq.~(\ref{eq:4}).
\item We will treat the environment-induced noise classically,
i.e., we replace the operator $\hat{\alpha}^{}_{z,j}$ by the classical random variable $\alpha^{}_{z,j}(t)$ 
(sometimes called a fluctuator) which
switches between $\pm1$. Accordingly, $\hat{\mathcal{X}}\rightarrow
\mathcal{X}(t)$ is the corresponding sum~(\ref{eq:4}) of such
random variables. Such a classical treatment is allowed provided
that the frequencies of the random transitions,
$\Gamma^{}_{1}\equiv T^{-1}_{1}$, are much smaller than the
temperature $T$ and the probed TLS energy splitting
$E$.\cite{GYM06}
\end{enumerate}

Having modelled the environment of TLSs as a classical noise, we
now consider the simplified model
\begin{align}
\label{eq:5}&\hat{H}=\frac{1}{2}E^{}_{p}\hat{\sigma}^{}_{z}+\frac{1}{2}\,X(t)\hat{\sigma}^{}_{z}\,
,
\end{align}
with
\begin{align}
\label{eq:6}X(t)&=\cos\theta^{}_{p}\,\mathcal{X}(t)=2\cos\theta^{}_{p}\sum_{j}J_{j}\cos\theta^{}_{j}\alpha^{}_{z,j}(t)\nonumber\\
&=\sum_{j}v^{}_{j}\alpha^{}_{z,j}(t)\, .
\end{align}
Here $v^{}_{j}$ is the effective coupling between the thermal TLS
at site $j$ and the probed TLS, given by
\begin{align}
\label{eq:7}&v^{}_{j}=2
J^{}_{j}\cos\theta^{}_{j}\cos\theta^{}_{p}\equiv 2
J^{}_{j}\sqrt{1-u^{}_{j}}\cos\theta^{}_{p}\, ,
\end{align}
with
$u^{}_{j}\equiv\sin^{2}\theta^{}_{j}=\left(\Delta{}_{0,j}/E^{}_{j}\right)^{2}$.
This model is often referred to as the spin-fluctuator
model.\cite{PE02,GYM04,BJ09} Let us now consider the effect of the
classical noise~(\ref{eq:6}), which consists of the environment of
thermal TLSs, on the Ramsey and echo decay signals.\cite{IG05}

The decay of a Ramsey signal is given by $f^{}_{R}(t)=\langle
e^{i\varphi_R(t)}\rangle$, where $\varphi_R(t)$ is the random
phase accumulated at time $t$,
\begin{align}
\label{eq:8}&\varphi_R(t)=-\int^{t}_{0}X(t')dt'\, .
\end{align}
In an echo experiment, the phase acquired is the difference
between the two free evolution periods,
\begin{align}
\label{eq:9}&\varphi^{}_{E}(t)=-\int^{t/2}_{0}X(t')dt'+\int^{t}_{t/2}X(t')dt'\,
,
\end{align}
and the corresponding decay function is $f^{}_{E}(t)=\langle
e^{i\varphi^{}_{E}(t)}\rangle$. Since we consider thermal TLSs
with $E^{}_{j}<T$, the transition rates in both directions can be
assumed to be equal. For a single fluctuator, characterized by
switching rate $\Gamma$ and coupling $v$, an average over the
switching history yields\cite{PE02,GYM04,BJ09}
\begin{align}
\label{eq:10}&f^{}_{R}(t)=e^{-\Gamma t}\left(\cos\mu t+\frac{\Gamma}{\mu}\sin\mu t\right),\nonumber\\
&f^{}_{E}(t)=e^{-\Gamma t}\left[1+\frac{\Gamma}{\mu}\sin\mu
t+\frac{\Gamma^{2}}{\mu^{2}}\left(1-\cos\mu t\right)\right]\, ,
\end{align}
where $\mu=\sqrt{v^{2}-\Gamma^{2}}$. The decay function due to a
set of fluctuators is the product of individual contributions of
the form~(\ref{eq:10}), that is
$F^{}_{R/E}(t)=\prod^{}_{j}f^{}_{R/E,j}(t)$.

It is useful to present the main limits of Eqs.~(\ref{eq:10}), as
shown in Table \ref{tbl}.
\begin{table}
\begin{tabular}{| c | c | c | c | c |}
\hline
& \multicolumn{2}{ |c| }{$v\gg\Gamma$} & \multicolumn{2}{ |c| }{$v\ll\Gamma$}\\
\hline
 & $vt\ll1$ & $vt \gg 1$ & $\Gamma t\ll 1$ & $\Gamma t\gg 1$ \\
 \hline
 $- \ln f^{}_{R} $ &  $\frac{v^2 t^2}{2}$ & $\, \Gamma t - \ln\left[\cos(vt)\right]$ & $\frac{v^2 t^2}{2}$ & $\frac{v^2}{\Gamma} t $ \\
 \hline
  $- \ln f^{}_{E} $ &  $\frac{v^2 \Gamma t^3}{6}$ & $\Gamma t$ & $\frac{v^2\Gamma t^3}{6}$ & $\frac{v^2}{\Gamma} t$\\
\hline
\end{tabular}
\vskip 0.3cm
\caption{\label{tbl} The different limits of the Ramsey and echo
decay signals, Eqs.~(\ref{eq:10}), due to a single fluctuator, characterized by
switching rate $\Gamma$ and coupling $v$.}
\end{table}
One can clearly see here the effect of the echo protocol. In the
short time limit, $vt \ll 1$ and $\Gamma t \ll 1$, the initial
parabolic decay $f^{}_{R}(T)\propto e^{-v^{2}t^{2}/2}$ is replaced
by the much slower decay $f^{}_{E}(t)\propto e^{-v^{2}\Gamma
t^{3}/6}$. In this regime, the decay due to a single fluctuator is small for both
Ramsey and echo protocols, but having many fluctuators might make this decay
relevant (see below).

For the regime $v>\Gamma$ and $vt\gg 1$ the oscillating entry
$\Gamma t-\ln\left[\cos(vt)\right]$ requires a separate
discussion. These oscillations were discovered in Ref.\
\onlinecite{GYM04} as an example of a prominent non-gaussian
effect. The origin of the oscillations is the interference between
two "paths" in which the equally probable $\pm$ state of the
fluctuator never changes during a single experimental run.
Therefore, these oscillations are almost completely removed by the
echo protocol, as can be seen in Table \ref{tbl}. If several
fluctuators in this regime contribute, the Ramsey decay function
will be a product of several incommensurate oscillating functions
and, thus, will be strongly suppressed. As we will see below, this
is the regime of the Ramsey decay, which occurs at times longer
than the typical dephasing times. Finally, in the regime
$v<\Gamma$ and $\Gamma t \gg 1$ the echo protocol is inefficient
and the result coincides with the one given by the Golden rule
(see more details in Sec.\ \ref{Sec 3C}).
\subsection{Averaging over the distribution of TLS parameters}
\label{Sec 1B} There are three parameters that control the thermal
TLSs: i) the energy
$E^{}_{j}=\sqrt{\Delta^{2}_{j}+\Delta^{2}_{0,j}}<T$, ii) the
normalized tunnel splitting
$u^{}_{j}\equiv\sin^{2}\theta^{}_{j}=\left(\Delta{}_{0,j}/E^{}_{j}\right)^{2}$,
and iii) the distance to the probed TLS $r^{}_{j}$. In a concrete
sample at a given strain we have a unique realization of the
parameters $\big\{E^{}_{j},u^{}_{j},r^{}_{j}\big\}$ characterizing
the TLSs, and the total Ramsey/echo decay signal is given by
\begin{align}
\label{eq:11}&-\ln F^{}_{R/E}(t)=-\sum^{}_{j}\ln
f^{}_{{R/E},j}(t)\, .
\end{align}
We first estimate $\langle \ln | F^{}_{R/E}(t)|\rangle_D$, where
$\langle \dots \rangle_D$ denotes the averaging over the disorder
configurations of the TLSs. Since we are dealing with quenched
disorder, the question of self-averaging arises and one should
examine if the result for $\langle \ln | F^{}_{R/E}(t)|\rangle_D$
corresponds to a typical situation.\cite{Schriefl} We obtain
\begin{eqnarray}
\label{eq:lnFgeneral} &&\langle \ln
|F^{}_{R/E}(t)|\rangle_D\nonumber\\&&=\int d\Omega
\int^{\infty}_{R^{}_{0}}
r^{D-1}dr\int^{1}_{u^{}_{\mathrm{min}}}du\int_0^{T}dE\,
P(E,u)\,\ln |f_{R/E}|\ ,\nonumber\\
\end{eqnarray}
where $D$ is the spatial dimension of the amorphous tunnel barrier
and
\begin{align}
\label{eq:13}&P(E,u)=\frac{P^{}_{0}}{2u\sqrt{1-u}}
\end{align}
is derived from the standard tunnelling model distribution
function $P(\Delta,\Delta^{}_{0})=P^{}_{0}/\Delta^{}_{0}$ using
the relations $E\equiv\sqrt{\Delta^{2}+\Delta^{2}_{0}}$ and
$u\equiv\sin^{2}\theta=\left(\Delta^{}_{0}/E\right)^{2}$ (see
Ref.~\onlinecite{HS88}). In Eq.~(\ref{eq:lnFgeneral}) $f_{R/E} =
f_{R/E}(v,\Gamma)$ are given by Eqs.~(\ref{eq:10}) and
$u^{}_{\mathrm{min}}$ is a lower cutoff for the parameter $u$. It
should be emphasized that $\Gamma(E,u)$ depends on $E$ and $u$
[see Eq.~(\ref{eq:18})], whereas $v(r,u)$ depends on $r$ and $u$
[see Eq.~(\ref{eq:7})].

\begin{figure}[ht!]
\begin{center}
\includegraphics[width=0.52\textwidth,height=0.23\textheight]{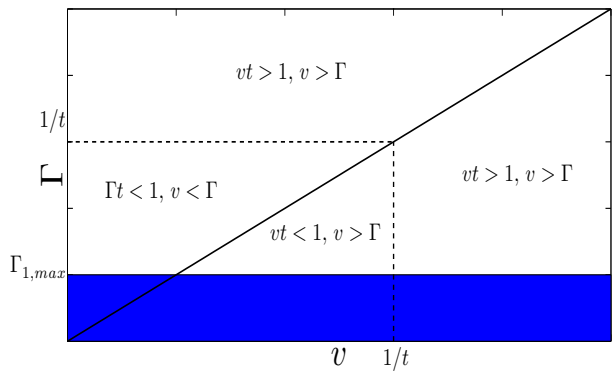}
\end{center}
\caption{\label{fig:Fig1}(Color online) The different domains in
the $v$-$\Gamma$ plane corresponding to the various limits of a
Ramsey and echo decay signals due to a single fluctuator, as
described in Table \ref{tbl}. The general
expressions~(\ref{eq:14}) and~(\ref{eq:15}) for the averaged
Ramsey and echo decay signals over an ensemble of TLSs are
obtained by dividing the integrations into the various domains and
setting the appropriate limits for $f^{}_{R/E}$ in each domain.
The shaded blue region shows the relevant domain for $\tau$-TLSs,
$\Gamma<\Gamma^{}_{1,\text{max}}\ll 1/t$.}
\end{figure}

According to Table \ref{tbl}, it is convenient to divide the
integration into three domains: a) $\Gamma t < 1$ and $v t <1$; b)
$v>\Gamma$ and $vt>1$; c) $\Gamma>v$ and $\Gamma t
>1$, as illustrated in Fig.\ \ref{fig:Fig1}. Thus, for the Ramsey decay signal we obtain
\begin{widetext}
\begin{align}
\label{eq:14}-\langle \ln |F^{}_{R}(t)|\rangle_D&\approx\int
d\Omega\int^{\infty}_{R^{}_{0}}
r^{D-1}dr\int^{1}_{u^{}_{\mathrm{min}}}du\int_0^{T}dE\,
P(E,u)\theta(1-vt)\theta(1-\Gamma t)\,\frac{v^{2}t^{2}}{2}\nonumber\\
&+\int d\Omega\int^{\infty}_{R^{}_{0}} r^{D-1}dr\int^{1}_{u^{}_{\mathrm{min}}}du\int^{T}_{0}dE\, P(E,u)\theta(vt-1)\theta(v-\Gamma)\,\left[\Gamma t - \ln|\cos(vt)| \right]\nonumber\\
&+\int d\Omega\int^{\infty}_{R^{}_{0}}
r^{D-1}dr\int^{1}_{u^{}_{\mathrm{min}}}du\int^{T}_{0}dE\, P(E,u)
\theta(\Gamma t-1)\theta(\Gamma-v)\,\frac{v^{2}t}{\Gamma}\, .
\end{align}
For the echo decay this would give
\begin{align}
\label{eq:15}-\langle \ln |F^{}_{E}(t)|\rangle_D&\approx\int
d\Omega\int^{\infty}_{R^{}_{0}}
r^{D-1}dr\int^{1}_{u^{}_{\mathrm{min}}}du\int^{T}_{0}dE\,
P(E,u)\theta(1-vt)\theta(1-\Gamma t)\,\frac{v^{2}\Gamma t^{3}}{6}\nonumber\\
&+\int d\Omega\int^{\infty}_{R^{}_{0}}
r^{D-1}dr\int^{1}_{u^{}_{\mathrm{min}}}du\int^{T}_{0}dE\,
P(E,u)\theta(vt-1)\theta(v-\Gamma)\,\Gamma t\nonumber\\
&+\int d\Omega\int^{\infty}_{R^{}_{0}}
r^{D-1}dr\int_{u^{}_{\mathrm{min}}}^{1}du\int_0^{T}dE\,
P(E,u)\theta(\Gamma t-1)\theta(\Gamma-v)\,\frac{v^{2}t}{\Gamma}\,
.
\end{align}
\end{widetext}
In the next section we use these expressions to study the average
Ramsey and echo decay signals that arise from the coupling of the
probed TLS to thermal $\tau$-TLSs. We show that for the Ramsey
dephasing there is no self-averaging, i.e.\ the average result
differs markedly from the typical one. For the echo dephasing, on
other hand, the result is self-averaging and the dephasing rate
calculated from Eq.~(\ref{eq:15}) coincides with the typical case.
\section{Dephasing by TLSs of the Standard Tunneling Model ($\tau$-TLSs)}
\label{Sec 2}
\subsection{Characteristics of $\tau$-TLSs}
\label{Sec 2A} According to Table \ref{tbl}, the effect of an
environment of thermal TLSs on the coherence properties of the
probed TLS depends on the typical couplings and relaxation (switching) rates
of the thermal TLSs. To estimate the relaxation rates of
$\tau$-TLSs, we assume that the random transitions of each TLS are
mainly due to its coupling to phonons. We will thus neglect
interactions between the TLSs comprising the noisy environment.
Additional relaxation mechanisms possible in superconducting
circuits, such as relaxation by quasiparticle excitations, are
negligible at $T<100\,$mK.\cite{Lisenfeld} Therefore, the rate of
random transitions of a single TLS is the relaxation rate due to
one-phonon processes,\cite{PWA72,JJ72}
\begin{align}
\label{eq:16}&\Gamma^{}_{1}=\frac{\left(2\pi\right)^{3}E\Delta^{2}_{0}\gamma^{2}}{\rho
h^{4}}\left(\frac{1}{v^{5}_{l}}+\frac{2}{v^{5}_{t}}\right)\coth(E/2T)\,
,
\end{align}
where $v^{}_{l}$ and $v^{}_{t}$ are the sound velocities of the
longitudinal and transverse modes, respectively. For a given
energy splitting $E$, the maximum relaxation rate is obtained for
$\Delta^{}_{0}=E$. Moreover, for thermal TLSs with $E\approx T$,
Eq.~(\ref{eq:16}) yields the maximum relaxation rate
\begin{align}
\label{eq:17}&\Gamma^{}_{1,\mathrm{max}}\approx2\frac{\left(2\pi\right)^{3}T^{3}\gamma^{2}}{\rho
h^{4}}\left(\frac{1}{v^{5}_{l}}+\frac{2}{v^{5}_{t}}\right)\, .
\end{align}
The relaxation rate~(\ref{eq:16}) for thermal TLSs ($E<T$) can be
written in terms of $E$ and $u$ as
\begin{align}
\label{eq:18}&\Gamma^{}_{1}(E,u)\approx\Gamma^{}_{1,\mathrm{max}}u\left(\frac{E}{T}\right)^{2}\,
,
\end{align}
where we replaced $\coth\left(E/2T\right)$ in Eq.~(\ref{eq:16}) by
$2T/E$ and used Eq.~(\ref{eq:17}). To estimate the maximum
relaxation rate for $\tau$-TLSs, we assume $\gamma\approx
1\,$eV.\cite{PWA87} Furthermore, since the dielectric layer
thickness is much smaller than the relevant phonon wavelength, the
sound velocity in Eqs.~(\ref{eq:16}) and~(\ref{eq:17}) is set by
the aluminum layers of the electrodes.\cite{SY10} At $T=35\,$mK
(the temperature corresponding to the measurements of Ref.\
\onlinecite{LJ15}) we obtain
\begin{align}
\label{eq:19}&\Gamma^{}_{1,\mathrm{max}}\approx
10\,\mathrm{ms}^{-1}\, ,
\end{align}
or equivalently, $T^{}_{1,\mathrm{min}}\approx 0.1\,$ms.

We next estimate the typical coupling between the probed TLS and
its nearest thermal TLS. The typical distance between the probed
TLS and its nearest thermal $\tau$-TLS, $R^{}_{T}$, can be
estimated using the relations
\begin{align}
\label{eq:20}&R^{3}_{T,3D}\int^{1}_{u^{}_{\mathrm{min}}}du\int^{T}_{0}dE\,P(E,u)=1
\;\: \left(\mathrm{3D}\right)\, ,\nonumber\\
&R^{2}_{T,2D}\,d\int^{1}_{u^{}_{\mathrm{min}}}du\int^{T}_{0}dE\,P(E,u)=1
\;\: \left(\mathrm{2D}\right)\, ,
\end{align}
where $u^{}_{\mathrm{min}}$ is a lower cutoff for the parameter
$u$ and $d$ is the tunnel barrier thickness ($d\approx 3\,$nm).
Here we provide the estimates in three and two dimensions (3D and
2D, respectively) due to the quasi-2D structure of the amorphous
tunnel barrier. In the framework of the STM, one usually assumes a
constant DOS, namely $P^{}_{0}$ in Eq.~(\ref{eq:13}) is a
constant. The central dimensionless parameter of the STM is the
tunneling strength $C^{}_{0}=P^{}_{0}\gamma^{2}/\rho
v^{2}=P^{}_{0}\,R^{3}_{0}\,J^{}_{0}\approx 10^{-3}$,\cite{PRO02}
where $R^{}_{0}$ and $J^{}_{0}$ are the typical distance and
typical interaction between nearest neighbors $\tau$-TLSs,
respectively.

Substituting Eq.~(\ref{eq:13}) into Eq.~(\ref{eq:20}), we readily
obtain $R^{3}_{T,3D}=R^{2}_{T,2D}\,d\approx 1/(P^{}_{0}\,\xi\,T)$
with $\xi=\ln\left(1/u^{}_{\mathrm{min}}\right)$. Inserting these
distances in the interaction coefficients [Eqs.~(\ref{eq:int2})
and~(\ref{eq:int3})], we find the typical coupling strength
between the probed TLS and its nearest thermal $\tau$-TLS,
\begin{align}
\label{eq:21}&J^{}_{T}=C^{}_{0}\,\xi\,T\sim 2\pi\cdot
10\,\mathrm{MHz} \;\:
\left(\mathrm{3D}\right)\, ,\nonumber\\
&J^{}_{T}=C^{}_{0}\,\xi\,T\left(\frac{d}{R^{}_{T,3D}}\right)^{3/2}\sim
2\pi\cdot 1\,\mathrm{MHz} \;\: \left(\mathrm{2D}\right)\, ,
\end{align}
where we assumed the usual values $C^{}_{0}\approx 10^{-3}$ and
$\xi\approx 20$. For the 2D estimate we also used $R^{}_{0}\approx
d/3$ and $J^{}_{0}\approx 1\,\text{K}\approx 2\pi\cdot 20\,$GHz.
As shown in Ref.\ \onlinecite{LJ15}, the 2D estimate of $J^{}_{T}$
is in good agreement with the observed magnitude of the Ramsey
dephasing rate (see more details below), whereas the 3D estimate
yields somewhat larger rates than observed experimentally. This
agrees with the quasi-2D structure of the junction barrier and
suggests that TLSs reside only in the amorphous tunnel barrier.

\begin{figure}[ht!]
\begin{center}
\includegraphics[width=0.4\textwidth,height=0.15\textheight]{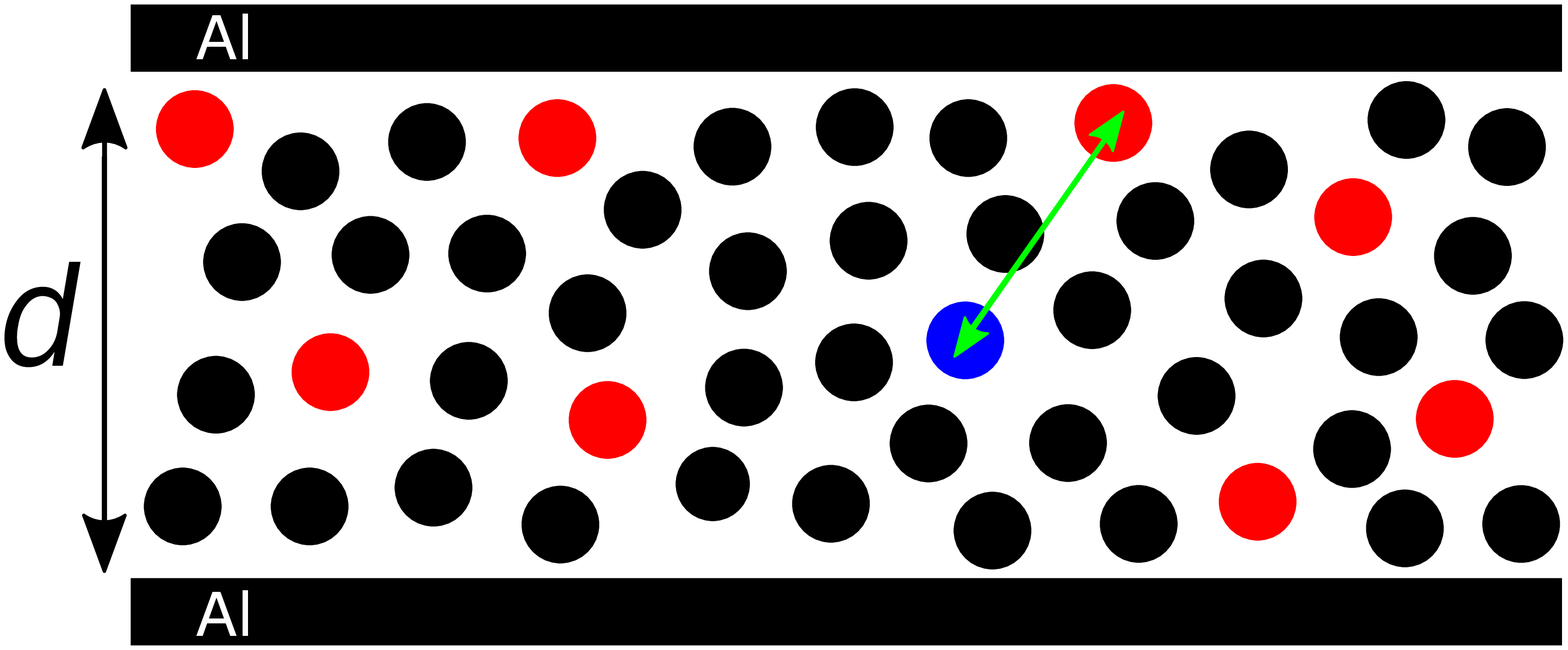}
\end{center}
\caption{\label{fig:Fig2}(Color online) Schematic view of TLSs
(circles) in the aluminum oxide tunnel barrier (of width $d$) of a
Josephson junction. The probed TLS is shown in blue and
non-thermal TLSs are shown in black. Thermal TLSs are denoted in
red and the typical distance $R^{}_{T}$ between the probed TLS and
its nearest thermal TLS is shown by a green arrow. Thermal TLSs
are typically in the regime $v\gg\Gamma$. As discussed in the
text, Ramsey dephasing is dominated by few thermal TLSs located at
$r\sim R^{}_{T}$ whereas echo dephasing is caused by a large
number of thermal TLSs.}
\end{figure}

Comparison of Eqs.~(\ref{eq:19}) and (\ref{eq:21}) reveals that
thermal $\tau$-TLSs satisfy $J^{}_{T}\gg\Gamma^{}_{1,\text{max}}$.
We recall Eq.~(\ref{eq:7}) and take into account that typically,
$\cos\theta^{}_{j}=O(1)$. Therefore, in a typical sample the most
significant thermal $\tau$-TLSs (i.e.\ those located at $r\sim
R^{}_{T}$) are characterized by $v\gg\Gamma$, except in the very
close vicinity of the symmetry point of the probed TLS,
$\Delta^{}_{p}=0$ (hence $\cos\theta^{}_{p}=0$), which can not be
distinguished within the resolution of $\Delta^{}_{p}$ in the
experiment. Therefore, in the case of $\tau$-TLSs we should study
the $\Delta^{}_{p}$-dependence of the dephasing rates assuming the
presence of fluctuators which are slow and strongly coupled to the
probed TLS, as sketched schematically in Fig.\ \ref{fig:Fig2}.
\subsection{Ramsey dephasing rate}
\label{Sec 2B} As explained above, the contribution of $\tau$-TLSs
to the Ramsey and echo decay signals should be treated within the
limit $v\gg\Gamma$. Under these circumstances the third line of
Eqs.~(\ref{eq:14}) and~(\ref{eq:15}) can be disregarded.
Furthermore, we examine now the regime of purely static dephasing,
$\Gamma_{1,\text{max}}t \ll 1$, which is the relevant regime for
the Ramsey dephasing studied in Ref.\ \onlinecite{LJ15}. Since the
observed dephasing times
$T^{}_{\varphi,R}=\Gamma^{-1}_{\varphi,R}$ are on the order of
microseconds, the condition $\Gamma_{1,\text{max}}t \ll 1$ indeed
holds at the relevant times $t<T^{}_{\varphi,R}$.

In this case the average Ramsey decay signal due to $\tau$-TLSs
can be estimated as
\begin{eqnarray}
\label{eq:22} &&\langle \ln
|F^{}_{R}(t)|\rangle_D\nonumber\\&&=\int
d\Omega\int^{\infty}_{R^{}_{0}}
r^{D-1}dr\int^{1}_{u^{}_{\mathrm{min}}}du\int^{T}_{0}dE\,P(E,u)\ln|\cos(vt)|\ ,\nonumber\\
\end{eqnarray}
where $v(r,u) = J^{}_{0}(R^{}_{0}/r)^{3}\sqrt{1-u}\cos\theta_p$
[see Eqs.~(\ref{eq:int2}), (\ref{eq:int3}), (\ref{eq:7})]. Using
Eq.~(\ref{eq:13}) and introducing an auxiliary variable $x\equiv t
J^{}_0 (R_0/r)^3\,|\cos\theta_p|$ we obtain for $D=3$
\begin{eqnarray}
\label{eq:23} &&\langle \ln
|F^{}_{R}(t)|\rangle_D=\frac{2\pi}{3}C^{}_{0}\,T t \,|
\cos\theta_p| \nonumber\\&& \times\int_{0}^{x_t}\frac{dx}{x^2}
\int^{1}_{u^{}_{\mathrm{min}}}du
\frac{\ln|\cos(x\sqrt{1-u})|}{u\sqrt{1-u}}\ ,
\end{eqnarray}
where $x^{}_{t}\equiv t J^{}_{0}|\cos\theta^{}_{p}|$. Setting
$J^{}_{0}\approx 1\,\text{K}\approx 2\pi\cdot 20\,$GHz, we observe
that at all relevant times (of the order of the dephasing time
$\Gamma^{-1}_{\varphi,R}$) $x_t \gg 1$, except in a very small
vicinity of the symmetry point $\cos\theta_p=0$. Thus we take the
limit $x^{}_{t}\rightarrow\infty$ in Eq.~(\ref{eq:23}). The
integral over $x$ converges, whereas integration over $u$ gives a
result of the order of
$\xi=\ln\left(1/u^{}_{\mathrm{min}}\right)$. Thus we obtain (for
$D=3$)
\begin{eqnarray}
\label{eq:lnFav3D} \langle \ln |F^{}_{R}(t)|\rangle_{D}  \sim -
J^{}_{T} |\cos \theta_p|\, t\ .
\end{eqnarray}
A similar calculation for $D=2$ gives
\begin{eqnarray}
\label{eq:lnFav2D} \langle \ln |F^{}_{R}(t)|\rangle_{D}  &\sim& -
\left[J^{}_T |\cos \theta_p|\right]^{2/3}\, t^{2/3}\ .
\end{eqnarray}
The coupling strength between the probed TLS and its closest
thermal TLS, $J^{}_{T}$, is defined in Eqs.~(\ref{eq:21}) for both
$D=3$ and $D=2$. The results (\ref{eq:lnFav3D}) and
(\ref{eq:lnFav2D}) turn out to be non-self-averaging.

\begin{figure}[ht!]
\begin{center}
\includegraphics[width=0.52\textwidth,height=0.23\textheight]{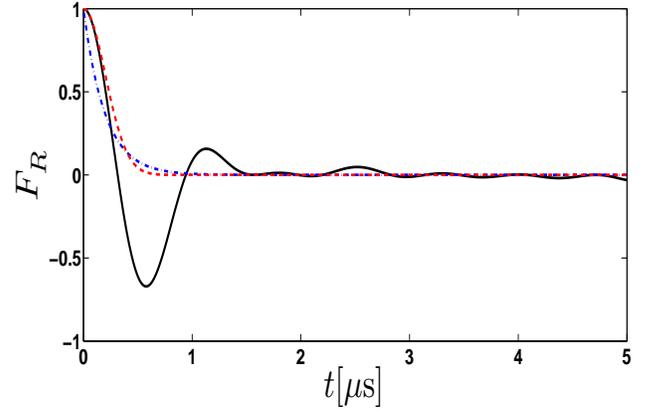}
\end{center}
\caption{\label{fig:Fig3}(Color online)  Comparison of
$e^{\langle\ln |F^{}_{R}(t)|\rangle^{}_{D}}$ obtained by averaging
over an ensemble of slow fluctuators with
$\Gamma\ll\Gamma^{}_{\varphi,R}$ as given by
Eq.~(\ref{eq:lnFav3D}) (dashed-dotted blue line), and of $F^{}_{R}(t)$ for a
specific realization of six thermal fluctuators with couplings
$v=2\pi\times\{5,1,0.8,0.5,0.4,0.2\}\,$MHz and relaxation rates
$\Gamma=\{5,0.5,8,0.8,0.1,3\}\,\mathrm{ms}^{-1}$ (solid black line).
At short times, $t\ll 1/v^{T}_{\text{max}}\sim 0.1\,\mu$s, the
Ramsey decay is well approximated by a typical Gaussian decay as
given in Eq.~(\ref{eq:lnFtyp}) (dashed red line).}
\end{figure}

Indeed, in a typical situation there is a closest thermal TLS with
maximum coupling $v^{T}_{\text{max}}\sim J^{}_{T}|\cos\theta_p|\gg
\Gamma_{1,\text{max}}$. At very short times, $t\ll
1/v^{T}_{\text{max}}$, one obtains $\ln |F^{}_{R}(t)|= -(1/2) t^2
\sum_j v_j^2$. Since $v_j \sim 1/r_j^3$ the sum is dominated by
the few closest thermal TLSs and in both $D=3$ and $D=2$
\begin{equation}
\label{eq:lnFtyp} \ln |F^{typ}_{R}(t)|  \sim - [J^{}_{T}\cos
\theta_p]^2\, t^2\ .
\end{equation}
Both the average [Eqs.~(\ref{eq:lnFav3D}) and~(\ref{eq:lnFav2D})]
and the typical [Eq.~\ref{eq:lnFtyp})] results give the dephasing
rate\cite{Comment2}
\begin{align}
\label{eq:25}&\Gamma^{}_{\varphi,R}\approx
J^{}_{T}|\cos\theta^{}_{p}|\, ,
\end{align}
yet with a very different functional time dependence. At short
times, $\Gamma^{}_{\varphi,R}\,t \ll 1$, the typical result gives
a much weaker Ramsey decay than would be naively expected from the
average result. At longer times, $\Gamma^{}_{\varphi,R}\,t \gg 1$,
the difference between the average and the typical results is even
more striking. The typical decay function $F^{typ}_{R}(t)$ at such
times oscillates between positive and negative
values,\cite{GYM04,BJ09} which means that the envelope of Ramsey
fringes has points of zero amplitude where phase slips occur. In
contrast, the average result is monotonically decaying. This
situation is demonstrated in Fig.~\ref{fig:Fig3}.

As demonstrated experimentally in Refs.\ \onlinecite{GGJ12} and
\onlinecite{LJ15}, the energy bias of the probed TLS changes
linearly with an external strain (which, in turn, is proportional
to a piezo voltage $V$). Therefore, close to the symmetry point
$\Delta^{}_{p}=0$, we can write
$\cos\theta^{}_{p}=\Delta^{}_{p}/E^{}_{p}$ as
\begin{align}
\label{eq:26}&\cos\theta^{}_{p}=\frac{\Delta^{}_{p}}{E^{}_{p}}=\frac{\Delta^{}_{p}}{\sqrt{\Delta^{2}_{p}+\Delta^{2}_{0,p}}}\approx\frac{\Delta^{}_{p}}{\Delta^{}_{0,p}}\,
.
\end{align}
Combining Eqs.~(\ref{eq:25}) and (\ref{eq:26}), one finds
$\Gamma^{}_{\varphi,R}\propto|\Delta^{}_{p}|$, i.e.\ the Ramsey
dephasing rate is expected to change linearly with
$\Delta^{}_{p}$. However, this linear dependence is expected to
hold as long as the Ramsey dephasing is dominated by the same
decohering thermal TLSs. As the strain is varied the asymmetry
bias $\Delta^{}_{j}$ of the decohering TLSs will change as well,
similarly to the probed TLS. As a result, at some point the energy
of a decohering TLS will exceed the thermal energy. At the same
time, other TLSs which were non-thermal at lower strains will
become thermal. Assuming that thermal TLSs are coupled to the
strain similarly as the probed TLS, one expects that once
$\Delta^{}_{p}$ is changed by $T\sim 2\pi\cdot 1\,$GHz, the
dominant decohering TLSs will be different ones. This gives rise
to a non-regular behavior, reflected in a change of slope or small
irregularities in the Ramsey dephasing rate as a function of
strain. Such features are indeed observed in the
$\Delta^{}_{p}$-dependence of the Ramsey dephasing rate\cite{LJ15}
and are strongly suggestive of the predominance of single thermal
TLSs in the Ramsey dephasing. As a result, we expect the pure
linear behavior to be relevant at a window
$|\Delta^{}_{p}|<2\pi\cdot 1\,$GHz around the symmetry point of
the probed TLS. Indeed, for the four TLSs studied in Ref.\
\onlinecite{LJ15}, the estimate for $J^{}_{T}$ in 2D is in excellent
agreement with the fitted slope of the linear line in the region
$|\Delta^{}_{p}|<2\pi\cdot 1\,$GHz.

Equations~(\ref{eq:21}) and (\ref{eq:25}) also allow us to predict
the temperature dependence (for low temperatures, $T<100\,$mK) of
the Ramsey dephasing rate. At a fixed strain, these equations
predict $\Gamma^{}_{\varphi,R}\propto T$ for the 3D case and
$\Gamma^{}_{\varphi,R}\propto T^{3/2}$ for the 2D case. For an
energy-dependent DOS of the form $n^{}_{\tau}(E)\propto
E^{\eta}$,\cite{BJ14,FL15,CA15} the corresponding temperature
dependence is $\Gamma^{}_{\varphi,R}\propto T^{1+\eta}$ and
$\Gamma^{}_{\varphi,R}\propto T^{3\left(1+\eta\right)/2}$, for 3D
and 2D, respectively.
\subsection{Echo dephasing rate}
\label{Sec 2C} To estimate the contribution of $\tau$-TLSs to the
ensemble average of the echo decay signal, we have to calculate
the first two integrals of Eq.~(\ref{eq:15}). For simplicity, we
perform the integration in 3D. The results in 2D are qualitatively
similar and given in appendix \ref{Sec appA}. The contribution to
the first integral comes from TLSs for which $vt<1$. Since for
$\tau$-TLSs $v\gg\Gamma$, the condition $\Gamma t<1$ is
automatically satisfied. Hence, the contribution to this integral
comes from TLSs which are located at distances $r>R^{}_{\ast}$
from the probed TLS, where
$R^{3}_{\ast}=2J^{}_{0}R^{3}_{0}\sqrt{1-u}|\cos\theta^{}_{p}|\,t$.
Therefore,
\begin{widetext}
\begin{align}
\label{eq:27}&\int^{\infty}_{R^{}_{0}}4\pi r^{2}dr\int^{1}_{0}du\int^{T}_{0}dE\,P(E,u)\theta(1-vt)\theta(1-\Gamma t)\,\frac{v^{2}\Gamma t^{3}}{6}\nonumber\\
&\approx\frac{2}{3}t^{3}\cos^{2}\theta^{}_{p}\,P^{}_{0}\left(J^{}_{0}R^{3}_{0}\right)^{2}\Gamma^{}_{1,\text{max}}\int^{\infty}_{R^{}_{\ast}}\frac{4\pi}{r^{4}}dr\int^{1}_{0}\frac{\sqrt{1-u}}{2}\,du\int^{T}_{0}\left(\frac{E}{T}\right)^{2}dE\nonumber\\
&=\frac{2\pi}{27}\,t^{2}|\cos\theta^{}_{p}|P^{}_{0}\,R^{3}_{0}\,J^{}_{0}\,T\,\Gamma^{}_{1,\text{max}}\approx\frac{2\pi}{27}\,t^{2}J^{}_{T}|\cos\theta^{}_{p}|\,\Gamma^{}_{1,\text{max}}\,\xi^{-1}\,
,
\end{align}
\end{widetext}
where $J^{}_{T}$ is given by the first of Eqs.~(\ref{eq:21}). The
main contribution to the second integral comes from the close
TLSs, for which $r<R^{}_{\ast}$. Again, the condition $v\gg\Gamma$
for $\tau$-TLSs ensures that the second step function is always
unity. Thus,
\begin{widetext}
\begin{align}
\label{eq:28}&\int^{\infty}_{R^{}_{0}}4\pi r^{2}dr\int^{1}_{0}du\int^{T}_{0}dE\,P(E,u)\theta(vt-1)\theta(v-\Gamma)\,\Gamma t\nonumber\\
&\approx tP^{}_{0}\,\Gamma^{}_{1,\text{max}}\int^{R^{}_{\ast}}_{0}4\pi r^{2}dr\int^{1}_{0}\frac{du}{2\sqrt{1-u}}\int^{T}_{0}\left(\frac{E}{T}\right)^{2}dE\nonumber\\
&=\frac{4\pi}{9}\,t^{2}|\cos\theta^{}_{p}|P^{}_{0}\,R^{3}_{0}\,J^{}_{0}\,T\,\Gamma^{}_{1,\text{max}}\approx\frac{4\pi}{9}\,t^{2}J^{}_{T}|\cos\theta^{}_{p}|\,\Gamma^{}_{1,\text{max}}\,\xi^{-1}\,
.
\end{align}
\end{widetext}
Neglecting a numerical prefactor of order unity, the sum of
Eqs.~(\ref{eq:27}) and~(\ref{eq:28}) gives
\begin{align}
\label{eq:29}\langle\ln
|F^{}_{E}(t)|\rangle^{}_{D}&\approx-t^{2}J^{}_{T}|\cos\theta^{}_{p}|\,\Gamma^{}_{1,\text{max}}\,\xi^{-1}\,
.
\end{align}
Let us check if this result is self-averaging. To
this end, we estimate the number of thermal TLSs, $N^{}_{\ast}$,
in a sphere of radius $R^{}_{\ast}$ at time
$t\approx\Gamma^{-1}_{\varphi,E}$, with $\Gamma^{-1}_{\varphi,E}$
the echo dephasing rate
\begin{align}
\label{eq:30}&\Gamma^{}_{\varphi,E}\approx\sqrt{\Gamma^{}_{1,\text{max}}\,\xi^{-1}\,J^{}_{T}|\cos\theta^{}_{p}|}\,
.
\end{align}
This yields
\begin{align}
\label{eq:31}N^{}_{\ast}&=\frac{4\pi}{3}\int^{1}_{u^{}_{\text{min}}}du\int^{T}_{0}dE\,P(E,u)R^{2}_{\ast}(t=\Gamma^{-1}_{\varphi,E})\nonumber\\
&\approx\frac{4\pi}{3}J^{}_{T}|\cos\theta^{}_{p}|\Gamma^{-1}_{\varphi,E}\approx\frac{4\pi}{3}\sqrt{\frac{\xi\,J^{}_{T}}{\Gamma^{}_{1,\mathrm{max}}}|\cos\theta^{}_{p}|}\nonumber\\
&\approx 1500\sqrt{|\cos\theta^{}_{p}|}\, .
\end{align}
Therefore, except for strains very close to the symmetry point
(which are outside the resolution of the experiment), a sphere of
radius $R^{}_{\ast}$ contains many thermal TLSs. The contribution
of $\tau$-TLSs to the echo decay signal is thus self-averaging,
and the echo dephasing rate is given by Eq.~(\ref{eq:30}). Since
thermal $\tau$-TLSs fluctuate very slowly, we expect the echo
dephasing rate to be much smaller than the Ramsey dephasing rate.
Using Eqs.~(\ref{eq:25}) and~(\ref{eq:30}), we find
\begin{align}
\label{eq:32}\frac{\Gamma^{}_{\varphi,R}}{\Gamma^{}_{\varphi,E}}=\sqrt{\frac{\xi\,J^{}_{T}}{\Gamma^{}_{1,\mathrm{max}}}|\cos\theta^{}_{p}|}\approx
400\sqrt{|\cos\theta^{}_{p}|}\, .
\end{align}
Thus, the echo protocol is very efficient in suppressing the noise
caused by fluctuations of thermal $\tau$-TLSs. This contradicts
the experimental observation, which shows that the echo dephasing
rate is much larger than the contribution of $\tau$-TLSs,
Eq.~(\ref{eq:30}). Furthermore, the $|\Delta^{}_{p}|^{1/2}$
dependence of the echo dephasing rate on the asymmetry bias,
predicted by Eq.~(\ref{eq:30}), is also in disagreement with the
experimental result, which shows a quadratic dependence. By
repeating the calculations in 2D one finds
$\Gamma^{}_{\varphi,E}\propto |\Delta^{}_{p}|^{2/5}$ (see appendix \ref{Sec appA}), which also
contradicts the experiment. In fact, among the three integrals of
Eq.~(\ref{eq:15}) only the third one can explain the quadratic
dependence of the echo dephasing rate. The contribution to this
integral comes from fluctuators with $\Gamma t>1$ and $\Gamma>v$,
namely from fast fluctuators that are weakly coupled to the probed
TLS. Such fluctuators are in the opposite regime of that
corresponding to $\tau$-TLSs. In the next section we show that
$S$-TLSs, proposed in Ref.\ \onlinecite{SM13}, obey the above
conditions and form a possible source of such a noise. We estimate
the order of magnitude of the echo dephasing rate due to such TLSs
and examine their relevance to the results of Ref.\
\onlinecite{LJ15}.
\section{Decoherence by TLSs strongly interacting with strain}
\label{Sec 3} 
The STM assumes that the coupling of TLSs to the
strain is unique or taken from a narrow distribution. Here we
consider the possibility that TLSs with much stronger coupling to
the strain generically exist in amorphous solids, as has recently been
proposed in Ref.\ \onlinecite{SM13}. We show that the existence of
a few or even a single thermal and strongly coupled to strain TLSs within the
amorphous barrier produces echo dephasing rate which is in
agreement with both the magnitude and the strain dependence found
experimentally. 
\subsection{The model with two types of TLSs}
\label{Sec 3A} The theory presented in Ref.\ \onlinecite{SM13}
suggests the existence of two types of TLSs, characterized by
different coupling strengths to strain fields. Weakly and strongly
coupled TLSs are characterized by coupling strength
$\gamma^{}_{\tau}$ and $\gamma^{}_{S}$ to strain fields,
respectively. The small dimensionless parameter of the theory is
then $g\equiv\gamma^{}_{\tau}/\gamma^{}_{S}$, which is expected \cite{SM13,GAA11,CA14a}
 to lie in the range $g\approx 0.01-0.03$.
The bimodality of the coupling strengths of TLSs to strain fields
also modifies the phonon-mediated interactions between the
TLSs.\cite{BJL77,BAL98,SM08} The Hamiltonian~(\ref{eq:int1}) is
then generalized to\cite{SM13,Comment3}
\begin{align}
\label{eq:1}&\hat{\mathcal{H}}^{}_{\tau
S}=\frac{1}{2}\sum^{}_{i\neq
j}\left(J^{\tau\tau}_{ij}\hat{\tau}^{}_{z,i}\hat{\tau}^{}_{z,j}+J^{\tau
S}_{ij}\hat{\tau}^{}_{z,i}\hat{S}^{}_{z,j}+J^{SS}_{ij}\hat{S}^{}_{z,i}\hat{S}^{}_{z,j}\right)\,
,
\end{align}
where $\hat{\tau}^{}_{z,i}$ and $\hat{S}^{}_{z,j}$ are the Pauli
matrices that represent the $\tau$- and $S$-TLSs at sites $i$ and
$j$, and the interaction coefficients are similar to those of
Eqs.~(\ref{eq:int2}) and~(\ref{eq:int3}),
\begin{align}
\label{eq:2}&J^{ab}_{ij}=\frac{c^{ab}_{ij}\gamma^{}_{a}\gamma^{}_{b}}{\rho
v^{2}R^{3}_{ij}}=c^{ab}_{ij}J^{ab}_{0}\left(\frac{R^{}_{0}}{R^{}_{ij}}\right)^{3}
\qquad a,b=\tau,S\, .
\end{align}
with $J^{ab}_{0}=\gamma^{}_{a}\gamma^{}_{b}/(\rho
v^{2}R^{3}_{0})$ and $R^{}_{0}$ being the distance between nearest
neighbor TLSs.

The implications of the interactions~(\ref{eq:1}) between the two
types of TLSs on their single particle density of states (DOS)
have been studied in detail in Refs.\ \onlinecite{CA14b} and
\onlinecite{CA14c}. It was found that a pseudo-gap exists in the
DOS of $S$-TLSs at low energies $E<J^{\tau S}_{0}=gJ^{SS}_{0}\sim
10\,{\rm K}\sim 2\pi \cdot 200\,$GHz. In contrast, the $\tau$-TLS DOS is well described by a
Gaussian with width $\sim J^{\tau S}_{0}$. Thus, at low energies
$E\ll J^{\tau S}_{0}\sim 10\,$K, $S$-TLSs are much more scarce
than $\tau$-TLSs. As a consequence, most low-temperature
properties, such as specific heat and thermal conductivity, are
dominated by the weakly interacting $\tau$-TLSs, which can be
identified as the well-known TLSs of the STM.
\subsection{Characteristics of $S$-TLSs}
\label{Sec 3B} We estimate the maximum relaxation rate of thermal
$S$-TLSs and their typical coupling to the probed $\tau$-TLS by
repeating the calculations of Sec.\ \ref{Sec 2A}. The relaxation
rate is given by Eqs.~(\ref{eq:16})-(\ref{eq:18}), with
$\gamma^{}_{\tau}$ replaced by
${\gamma^{}_{S}=\gamma^{}_{\tau}/g}$. This yields the maximum
relaxation rate
\begin{align}
\label{eq:33}&\Gamma^{(S)}_{1,\text{max}}=\frac{\Gamma^{(\tau)}_{1,\text{max}}}{g^{2}}\approx
10-100\,\mu\mathrm{s}^{-1}\, ,
\end{align}
where we used Eq.~(\ref{eq:19}) and $g\approx 0.01-0.03$.

\begin{figure}[ht!]
\begin{center}
\includegraphics[width=0.4\textwidth,height=0.15\textheight]{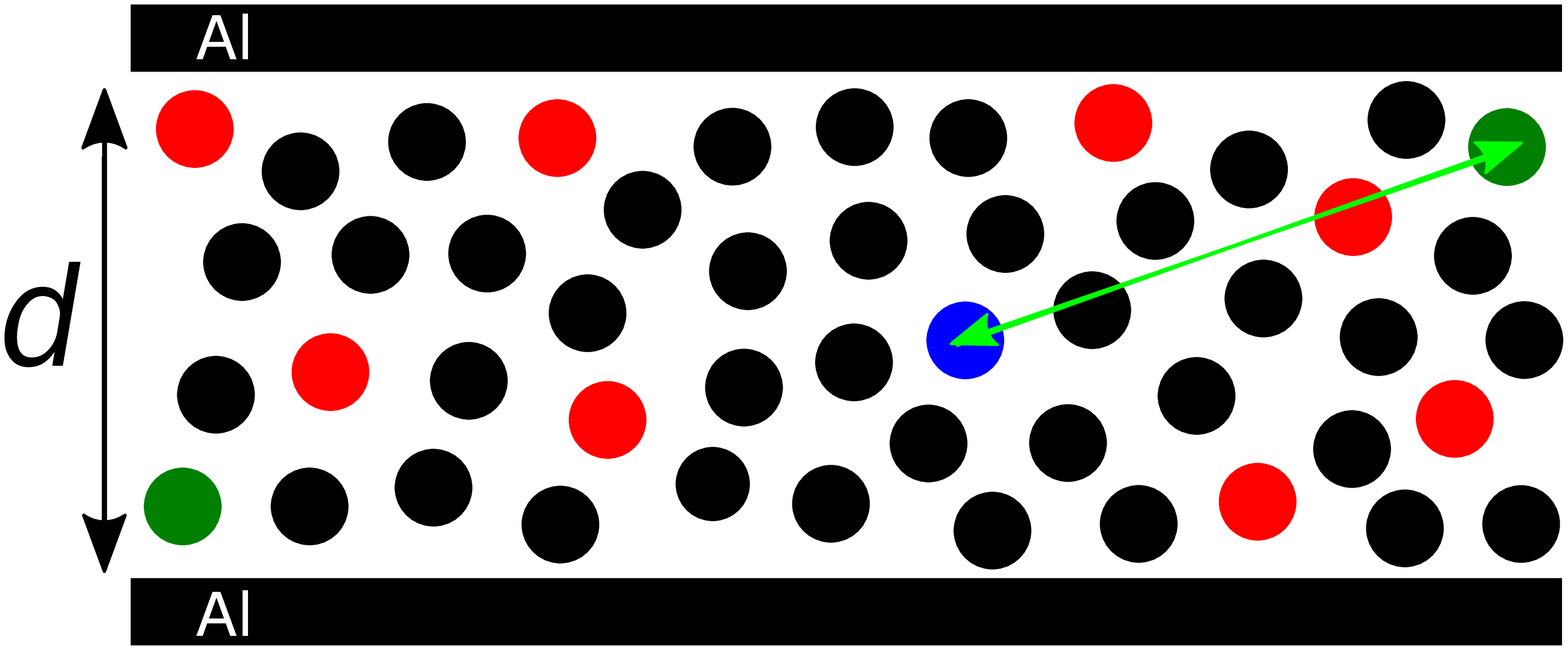}
\end{center}
\caption{\label{fig:Fig4}(Color online) Schematic view of a bath
of thermal $S$-TLSs (green circles) and $\tau$-TLSs (red circles)
coupled to the probed TLS (blue circle). Non-thermal TLSs are
shown in black circles. Thermal $S$-TLSs are much more scarce than
thermal $\tau$-TLSs and thus located much farther from the probed
TLS (the typical distance $R^{}_{T}$ is shown by a green arrow).
Thus, in contrast to $\tau$-TLSs, $S$-TLSs are fast and weakly
coupled to the probed TLS, i.e.\ they are typically in the regime
$\Gamma t>1$ and $v<\Gamma$.}
\end{figure}

In appendix \ref{Sec appB} we derive the typical distance between
the probed TLS and its nearest thermal $S$-TLS and conclude that it is
somewhat larger than the size of the amorphous tunnel barrier used
in Ref.\ \onlinecite{LJ15} ($\sim100\,$nm). We therefore assume
the existence of one or few $S$-TLSs in the amorphous layer,
relying also on the possibility of an enhanced $S$-TLSs DOS on the
surface.\cite{FL12} Thus, the typical distance between the probed TLS and its
nearest thermal $S$-TLS is of the order of the amorphous layer
size which gives a typical interaction
\begin{align}
\label{eq:33b}&J^{\tau S}_{T}\approx 2\pi\times (1-10)\,\mathrm{MHz}\, .
\end{align}
Therefore, one finds that for $S$-TLSs $J^{\tau
S}_{T}<\Gamma^{(S)}_{1,\text{max}}$. It thus follows that in a
typical sample the thermal $S$-TLSs are characterized by $v<\Gamma$,
as shown schematically in Fig.\ \ref{fig:Fig4} (cf.\ Fig.\
\ref{fig:Fig2}). As argued above, such fluctuators give rise to a
quadratic dependence of the echo dephasing rate on the asymmetry
bias $\Delta^{}_{p}$. We now study in more detail the effects of
$S$-TLSs on the Ramsey and echo decay signals.
\subsection{Ramsey and Echo dephasing rates due to $S$-TLSs}
\label{Sec 3C} For $S$-TLSs, which satisfy $v<\Gamma$, we neglect
the second integral of Eqs.~(\ref{eq:14}) and~(\ref{eq:15}). At
short times, $t<1/\Gamma^{(S)}_{\text{max}}$, only the first
integral of Eqs.~(\ref{eq:14}) and~(\ref{eq:15}) contributes.
Similarly to Eq.~(\ref{eq:lnFtyp}), the typical Ramsey decay
signal is governed by the nearest thermal $S$-TLSs, i.e.\
\begin{align}
\label{eq:36}&\ln F^{typ}_{R}(t)\approx -\left[J^{\tau
S}_{T}\cos\theta^{}_{p}\right]^{2}t^{2}\, .
\end{align}
This quantity is very small for $t<1/\Gamma^{(S)}_{1,\text{max}}$,
so dephasing has not yet occurred at such short times. The typical
echo decay signal is of the order
\begin{align}
\label{eq:37}&\ln F^{typ}_{E}(t)\approx -\left[J^{\tau
S}_{T}\cos\theta^{}_{p}\right]^{2}\Gamma^{(S)}_{1,\text{max}}t^{3}\,
,
\end{align}
which is also very small. It should also be emphasized that in
contrast to $\tau$-TLSs, the contribution of $S$-TLSs to the
average echo decay signal is not self-averaging, since the first
integral of Eq.~(\ref{eq:15}) is dominated by rare samples in
which the nearest thermal $S$-TLS is located at $r\sim R^{}_{0}$.

As $t$ becomes longer than $1/\Gamma^{(S)}_{1,\text{max}}$, both
Ramsey and echo decay signals are dominated by the third integral
of Eqs.~(\ref{eq:14}) and~(\ref{eq:15}). Since this integral also
involves $v^{2}$, the main contribution comes from the thermal
$S$-TLS for which the quantity $v^{2}/\Gamma$ is maximum. Thus, in
a typical sample we expect the order of magnitude of the
contribution of $S$-TLSs to the Ramsey and echo decay signals to
be
\begin{align}
\label{eq:38}&\ln F^{typ}_{R}(t)=\ln F^{typ}_{E}(t)\approx
-\frac{\left[J^{\tau
S}_{T}\cos\theta^{}_{p}\right]^{2}}{\Gamma^{(S)}_{1,\text{max}}}t\,
.
\end{align}
The Ramsey and echo dephasing rates in 2D are thus
\begin{align}
\label{eq:39}\Gamma^{}_{\varphi,R}&=\Gamma^{}_{\varphi,E}\approx\frac{\left(J^{\tau
S}_{T}\right)^{2}}{\Gamma^{(S)}_{1,\text{max}}}\cos^{2}\theta^{}_{p}\nonumber\\
&\approx2\pi\times\cos^{2}\theta^{}_{p}\,\text{MHz}\, ,
\end{align}
where in the last step we used Eqs.~(\ref{eq:33})
and~(\ref{eq:33b}). Close to the symmetry point Eq.~(\ref{eq:39})
predicts a quadratic dependence of the echo dephasing rate on the
asymmetry bias, in agreement with the experiment. Moreover, the
order of magnitude of the echo dephasing rate is also in line with
the experimental result. We note that the above result holds
provided that at least one $S$-TLS exists within the amorphous
layer. Otherwise, we expect the echo dephasing to be negligible.
This is indeed observed in one out of four TLSs in the experiment
in Ref.\ \onlinecite{LJ15}.

The result~(\ref{eq:39}) for the Ramsey and echo dephasing rates
due to fast fluctuators can also be understood in terms of the
Gaussian approximation, since in the regime $\Gamma>v$ the results
of the spin-fluctutator model are expected to coincide with the
Gaussian approximation.\cite{BJ09} Within the Gaussian
approximation, where $X(t)$ in Eqs.~(\ref{eq:6}), (\ref{eq:8})
and~(\ref{eq:9}) is assumed to have a Gaussian statistics, one
ends up with
\begin{align}
\label{eq:40}&-\ln
F^{G}_{R/E}(t)=\frac{t^{2}}{2}\int^{\infty}_{-\infty}\frac{d\omega}{2\pi}S^{}_{X}(\omega)K^{}_{R/E}(\omega,t)\, ,\nonumber\\
&K^{}_{R}(\omega,t)=\sinc^{2}\left(\frac{\omega t}{2}\right)\, ,\nonumber\\
&K^{}_{E}(\omega,t)=\sin^{2}\left(\frac{\omega
t}{4}\right)\sinc^{2}\left(\frac{\omega t}{4}\right)\, ,
\end{align}
where $\sinc(x)\equiv\sin(x)/x$ and $S^{}_{X}(\omega)$ is the
noise spectral density, i.e.\ the Fourier transform of the
correlation function $S^{}_{X}(t)=\langle X(t)X(0)\rangle$. If
$S^{}_{X}(\omega)$ is approximately constant for
$|\omega|\lesssim\Gamma^{}_{\varphi,E/R}$,\cite{Comment4} namely
$X(t)$ is a white noise, then $-\ln
F^{G}_{R/E}(t)=\Gamma^{}_{\varphi,R/E}t$ where
\begin{align}
\label{eq:41}&\Gamma^{}_{\varphi,R}=\Gamma^{}_{\varphi,E}\approx\frac{1}{2}S^{}_{X}(\omega=0)\,
.
\end{align}
Using Eq.~(\ref{eq:6}) and assuming independent fluctuators, one
finds\cite{BJ09}
\begin{align}
\label{eq:42}S^{}_{X}(\omega)=\sum^{}_{j}v^{2}_{j}S^{}_{j}(\omega)=\cos^{2}\theta^{}_{p}\sum^{}_{j}J^{2}_{j}\cos^{2}\theta^{}_{j}S^{}_{j}(\omega)\,
,
\end{align}
where $S^{}_{j}(\omega)$ is the noise spectral density of a single
fluctuator,
\begin{align}
\label{eq:43}S^{}_{j}(\omega)=\frac{2\Gamma^{}_{j}}{\Gamma^{2}_{j}+\omega^{2}}\,
.
\end{align}
Equation~(\ref{eq:41}) then yields
\begin{align}
\label{eq:44}&\Gamma^{}_{\varphi,R}=\Gamma^{}_{\varphi,E}\approx\sum^{}_{j}\frac{J^{2}_{j}\cos^{2}\theta^{}_{j}}{\Gamma^{}_{j}}\cos^{2}\theta^{}_{p}\,
.
\end{align}
For an interaction which falls off as $1/R^{3}$, the sum is
dominated by a few closet fluctuators and the result~(\ref{eq:44})
is of the same order of magnitude as~(\ref{eq:39}).
\section{Discussion and conclusions}
\label{Sec 5} We have investigated the strain-dependent dephasing rates
of individual TLSs in the amorphous tunnel barrier of a
superconducting phase qubit, as recently studied in Ref.\
\onlinecite{LJ15}. The interaction between the probed TLS and
thermal standard ($\tau$-) TLSs described in the framework of the
STM is capable of explaining the observed Ramsey dephasing rate. This includes:
\begin{enumerate}[(1)]
\item The linear dependence on the applied strain and hence on the
bias energy of the probed TLS,
$\Gamma^{}_{\varphi,R}\propto|\Delta^{}_{p}|$, near the symmetry
point, $\Delta^{}_{p}=0$. It is a consequence
of the small relaxation rate of thermal $\tau$-TLSs compared to
their interaction with the probed TLS.
\item The deviation from linear behavior and the appearance of
irregularities when $|\Delta^{}_{p}|$ is
varied by an amount of the order of the thermal energy $T\approx 2\pi\cdot 1\,$GHz. Such
features are expected since the theory predicts the Ramsey
dephasing to be dominantly caused by a small number of TLSs. As
the strain is varied, some of these TLSs are no longer thermal
while other TLSs will become so and contribute to the Ramsey
dephasing.
\item The magnitude of the Ramsey dephasing rate. Using standard
estimates for the coupling of $\tau$-TLSs to strain fields,
$\gamma\approx 1\,$eV, and for the dimensionless "tunneling
strength" $C^{}_{0}=P^{}_{0}\gamma^{2}/\rho v^{2}\approx 10^{-3}$,
it is shown that the order of magnitude of the
Ramsey dephasing rate is in excellent agreement with the
experimental result. The estimate for a quasi-2D configuration is
in better agreement than that in a 3D configuration which is
consistent with the experimental setup. Although we considered
elastic interactions between $\tau$-TLSs, it should be noted that
our analysis and results can be carried through to the case where
electric dipole interactions are present between $\tau$-TLSs. For
typical values of the dipole moment, of the order of
$5\,$Debye,\cite{MJM05,PAP14,KMS14,BAL15} electric dipole interactions
are expected to be of the same order of magnitude as elastic
interactions between $\tau$-TLSs and the results are the same.
\end{enumerate}
Furthermore, we predict the dependence of the Ramsey dephasing rate on
temperature for both 2D and 3D geometries for an arbitrary
power-law energy dependence of the TLS DOS at low energies.

For the echo dephasing, our analysis shows that the contribution of
$\tau$-TLSs is very much reduced
($\Gamma^{}_{\varphi,E}/\Gamma^{}_{\varphi,R}\sim 0.01$ for the
typical strains studied in Ref.\ \onlinecite{LJ15}) and the
bias-dependence is predicted to be
$\Gamma^{}_{\varphi,E}\propto|\Delta^{}_{p}|^{0.4}$
(or $\Gamma^{}_{\varphi,E}\propto|\Delta^{}_{p}|^{0.5}$ in 3D). Such
small dephasing rates are observed only in one out of four TLSs
studied in Ref.\ \onlinecite{LJ15}. In the other TLSs the echo
dephasing rate is higher and varies quadratically with the
applied strain, implying the presence of energy fluctuations on a
time scale $\lesssim 1\,\mu\rm{s}$. We suggest that TLSs that are strongly
coupled to strain\cite{SM13} and therefore fast fluctuating may be the dominant source for
noise and governing echo dephasing at low temperatures. The properties of such TLSs, in particular their scarcity compared to the standard TLSs of the STM, result in both the order of magnitude and the strong fluctuations of the echo dephasing as seen in the experiment \cite{LJ15}. 
\begin{acknowledgments}
We acknowledge fruitful discussions with J. Lisenfeld. This work
was supported by the German-Israeli Foundation (GIF) and by the DFG Research Grant SCHO 287/7-1, SH 81/2-1.
\end{acknowledgments}
\appendix
\section{Calculation of the echo dephasing rate due to $\tau-$TLSs in
2D} \label{Sec appA} We give here the results of the calculations
performed in Sec.\ \ref{Sec 2C} for the 2D case. The integrals
\ref{eq:27} and \ref{eq:28} in 2D give
\begin{widetext}
\begin{align}
\label{eq:appA_eq1}&d\int^{\infty}_{R^{}_{0}}2\pi
rdr\int^{1}_{0}du\int^{T}_{0}dE\,P(E,u)\theta(1-vt)\theta(1-\Gamma
t)\,\frac{v^{2}\Gamma t^{3}}{6}\approx\frac{\pi d}{120\,\xi
R^{}_{0}}\left(\frac{J^{}_{T,3D}}{J^{}_{0}}\right)^{1/3}\,t^{5/3}\left[J^{}_{T,3D}|\cos\theta^{}_{p}|\right]^{2/3}\Gamma^{}_{1,\text{max}}\,
,
\end{align}
\begin{align}
\label{eq:appA_eq2}&d\int^{\infty}_{R^{}_{0}}2\pi
rdr\int^{1}_{0}du\int^{T}_{0}dE\,P(E,u)\theta(vt-1)\theta(v-\Gamma)\,\Gamma
t\approx\frac{2\pi d}{5\,\xi
R^{}_{0}}\left(\frac{J^{}_{T,3D}}{J^{}_{0}}\right)^{1/3}\,t^{5/3}\left[J^{}_{T,3D}|\cos\theta^{}_{p}|\right]^{2/3}\Gamma^{}_{1,\text{max}}\,
,
\end{align}
\end{widetext}
where $J^{}_{T,3D}$ is given by the first of Eqs.~(\ref{eq:21}).
The sum of Eqs.~(\ref{eq:appA_eq1}) and~(\ref{eq:appA_eq2}) yields
[cf.\ Eq.~(\ref{eq:29})]
\begin{align}
\label{eq:appA_eq3}\langle\ln
|F^{}_{E}(t)|\rangle^{}_{D}&\approx-\frac{d}{\xi
R^{}_{0}}\left(\frac{J^{}_{T,3D}}{J^{}_{0}}\right)^{1/3}\,t^{5/3}\left[J^{}_{T,3D}|\cos\theta^{}_{p}|\right]^{2/3}\nonumber\\
&\times\Gamma^{}_{1,\text{max}}\, ,
\end{align}
where a numerical coefficient of order unity is omitted. The echo
dephasing rate is thus
\begin{align}
\label{eq:appA_eq4}&\Gamma^{}_{\varphi,E}\approx\left[\frac{J^{}_{T,3D}\,\Gamma^{}_{1,\text{max}}d}{\xi
R^{}_{0}\left(J^{}_{0}\right)^{1/3}}\right]^{3/5}|\cos\theta^{}_{p}|^{2/5}\,
.
\end{align}
Equation~(\ref{eq:31}) takes the form
\begin{align}
\label{eq:appA_eq5}N^{}_{\ast}&=\pi
d\int^{1}_{u^{}_{\text{min}}}du\int^{T}_{0}dE\,P(E,u)R^{2}_{\ast}(t=\Gamma^{-1}_{\varphi,E})\nonumber\\
&\approx\frac{\pi d}{R^{}_{0}}\left(\frac{J^{}_{T,3D}}{J^{}_{0}}\right)^{1/3}\left[\frac{J^{}_{T,3D}|\cos\theta^{}_{p}|}{\Gamma^{}_{\varphi,E}}\right]^{2/3}\nonumber\\
&\approx\pi\left[\frac{d^{3/2}\xi J^{}_{T,3D}}{R^{3/2}_{0}\Gamma^{}_{1,\mathrm{max}}}\sqrt{\frac{J^{}_{T,3D}}{J^{}_{0}}}|\cos\theta^{}_{p}|\right]^{2/5}\nonumber\\
&\approx 100|\cos\theta^{}_{p}|^{2/5}\, .
\end{align}
Therefore, except for strains very close to the symmetry point
(which are outside the resolution of the experiment), a circle of
radius $R^{}_{\ast}$ contains many thermal TLSs. The contribution
of $\tau$-TLSs to the echo decay signal in 2D is thus also
self-averaging, and the echo dephasing rate is given by
Eq.~(\ref{eq:appA_eq4}). The ratio between the echo dephasing rate
and the Ramsey dephasing rate is
\begin{align}
\label{eq:appA_eq6}\frac{\Gamma^{}_{\varphi,R}}{\Gamma^{}_{\varphi,E}}&=\left(\frac{\xi
J^{}_{T,3D}|\cos\theta^{}_{p}|}{\Gamma^{}_{1,\text{max}}}\right)^{3/5}\left(\frac{d}{R^{}_{0}}\right)^{9/10}\left(\frac{J^{}_{T,3D}}{J^{}_{0}}\right)^{3/10}\nonumber\\
&\approx 300|\cos\theta^{}_{p}|^{3/5}\, ,
\end{align}
which suggests that the echo protocol is very efficient also in
2D.
\section{Density of thermal strongly interacting TLSs} \label{Sec appB}
To estimate the typical distance $R^{}_{T}$ between the probed TLS
and its nearest thermal $S$-TLS, we use the relations [cf.\
Eqs.~(\ref{eq:20})]
\begin{align}
\label{eq:34}&R^{3}_{T,3D}\int^{T}_{0}dE\,n^{}_{S}(E)=1 \;\: \left(\mathrm{3D}\right)\, ,\nonumber\\
&R^{2}_{T,2D}\,d\int^{T}_{0}dE\,n^{}_{S}(E)=1 \;\:
\left(\mathrm{2D}\right)\, ,
\end{align}
where $n^{}_{S}(E)$ is the DOS of $S$-TLSs. For $E<J^{\tau
S}_{0}$, the interaction between the TLSs [Eqs.~(\ref{eq:1})
and~(\ref{eq:2})] gives rise to a power-low pseudo-gap in the DOS
of $S$-TLSs, namely $n^{}_{S}(E)\propto
E^{\mu}$.\cite{SM13,CA14a,CA14b} Moreover, the DOS of $S$-TLSs is
related to that of $\tau$-TLSs via
$n^{}_{S}(E)=g^{2}n^{}_{\tau}(E)$ at $E=T^{}_{U}$, where
$T^{}_{U}\approx 1\,$K is the temperature below which universality
in amorphous and disordered systems is observed.\cite{SM13} Since
the DOS of $\tau$-TLSs is assumed to be constant,
$n^{}_{\tau}(E)\approx P^{}_{0}\,\xi$, it follows that
$n^{}_{S}(E)\approx
g^{2}P^{}_{0}\,\xi\left(E/T^{}_{U}\right)^{\mu}$. Using
Eqs.~(\ref{eq:34}), we readily obtain
\begin{align}
\label{eq:34b}R^{3}_{T,3D}&=R^{2}_{T,2D}\,d=\left(\mu+1\right)T^{\mu}_{U}/\left(g^{2}P^{}_{0}\,\xi\,T^{\mu+1}\right)\nonumber\\
&=\frac{\mu+1}{g^{2}C^{}_{0}\xi}\left(\frac{T^{}_{U}}{T}\right)^{\mu}\frac{J^{\tau\tau}_{0}}{T}R^{3}_{0}\,
.
\end{align}
Setting $T\approx 35\,$mK, $T^{}_{U}\approx 1\,$K and
$J^{\tau\tau}_{0}\approx 1\,$K, we obtain
$R^{}_{T,2D}\approx10^{3}R^{}_{0}$ (we note that this result is
weakly dependent on $\mu$). Assuming $R^{}_{0}\approx 1-2\,$nm we
find $R^{}_{T,2D}\sim 1\,\mu$m, somewhat larger than the sample
size. For completeness we give here the typical coupling at this
distance. Inserting Eq.~(\ref{eq:34b}) in the interaction
coefficients~(\ref{eq:2}), we find the typical coupling strength
between the probed TLS and its nearestc thermal $S$-TLS,
\begin{align}
\label{eq:35}J^{\tau
S}_{T}&=\frac{g}{\mu+1}\left(\frac{T}{T^{}_{U}}\right)^{\mu}J^{\tau\tau}_{T,3D}\sim
2\pi\cdot 10g\,\text{MHz} \;\: \left(\mathrm{3D}\right)\, ,\nonumber\\
J^{\tau
S}_{T}&=g^{2}\left[\frac{T^{\mu}}{\left(\mu+1\right)T^{\mu}_{U}}\right]^{3/2}J^{\tau\tau}_{T,2D}\sim
2\pi\cdot g^{2}\,\mathrm{MHz} \;\: \left(\mathrm{2D}\right)\, ,
\end{align}
where the coupling strength between the probed TLS and its nearest
thermal $\tau$-TLS, $J^{\tau\tau}_{T}$, is defined in
Eqs.~(\ref{eq:21}) for both $D=3$ and $D=2$.


\begin{references}
\bibitem{Strain} As mentioned above, the strain field in the absence of any external perturbation is also responsible for the relaxation of the probed TLS, discussed in Ref.\ \onlinecite{LJ15}, and for the TLS-TLS interactions [see Eqs.~(\ref{eq:int1})-(\ref{eq:int3}) and the third term of Eq.~(\ref{eq:Hpj})]. In Eq.~(\ref{eq:ProbedH}) we focus only on the applied strain $\epsilon_p$, whose sole effect is to externally tune the energy bias of the probed TLS.
\bibitem{Comment1} We set $\hbar=k^{}_{B}=1$ and measure energies in units of frequencies or temperatures.
\bibitem{Comment2} We use the term "rate" to refer to the inverse of the time for which the signal has decayed to $e^{-1}$. The decay, however, might not be an exponential one.
\bibitem{Comment3} The low-energy effective Hamiltonian includes also a random field term of the form $\sum^{}_{i}\left(h^{\tau}_{i}\hat{\tau}^{z}_{i}+h^{S}_{i}\hat{S}^{z}_{i}\right)$, where the random fields $h^{\tau}_{}$ and $h^{S}_{}$ have typical strengths $h^{\tau}_{}\lesssim gJ^{}_{0}$ and $h^{S}_{}\lesssim J^{}_{0}$. Such a term has no effect on the dynamics of the TLSs and can be absorbed into the energy bias $\Delta$.
\bibitem{Comment4} This condition is equivalent to $\tau^{}_{c}\lesssim T^{}_{\varphi,E/R}$ where $\tau^{}_{c}$ is the time over which the correlation function $S^{}_{X}(t)$ decays.
\end{references}
\end{document}